# Probabilistic Estimation of Software Project Duration

**A.M. Connor**
*Auckland University of Technology*
andrew.connor@aut.ac.nz

**Abstract**
*This paper presents a framework for the representation of uncertainty in the estimates for software design projects for use throughout the entire project lifecycle. The framework is flexible in order to accommodate uncertainty in the project and utilises Monte Carlo simulation to compute the propagation of uncertainty in effort estimates towards the total project uncertainty and therefore gives a project manager the means to make informed decisions throughout the project life. The framework also provides a mechanism for accumulating project knowledge through the use of a historical database, allowing effort estimates to be informed by, or indeed based upon, the outcome of previous projects. Initial results using simulated data are presented and avenues for further work are discussed.*

**Introduction**

Estimation of cost and duration for software development activities is one of the most difficult aspects of software project management. The project manager often has the responsibility to make accurate estimations of effort and cost against which a project's success will be judged. This is particularly true for projects subject to competitive bidding where a high bid could result in losing the contract or a low bid could result in a loss to the organisation. From an estimate, the management often decides whether to proceed with the project. Industry has a need for accurate estimates of effort and size at a very early stage in a project.

This paper, which extends an earlier conference paper (Connor & MacDonell, 2006), outlines a methodology for introducing probabilistic modelling for the estimation of duration for software development projects. Software development, more so than many other disciplines, is plagued by vague or shifting requirements and a lack of understanding regarding product complexity that often leads to projects being delivered either late, over budget or not to requirements. Software cost estimates made early in the software development process are often based on wrong or incomplete requirements.

In this paper, uncertainty in effort estimates are linked to a project work breakdown structure in an effort to achieve two purposes. Initially, the method described in this paper can be utilised during the development of a tender submission for a software project. Typically in this circumstance, a project effort estimate will be made using a number of methods, such as expert opinion or a parametric model. It is possible that some companies will base a bid/no-bid decision on this single estimate without any deeper analysis of the risks involved. The tool detailed in this paper, developed in Excel using a freely available add-in, *SimulAr* (Machain, 2005), allows uncertainty in estimates to be captured and use of Monte-Carlo simulation provides an indication of the range of likely outcomes, not just a single estimate. The bid/no-bid decision can therefore be informed by pessimistic, optimistic and realistic estimates. In addition, the analysis of the data allows the areas of highest risk to be located and as such the project manager can allocate resource in the development of the tender submission in order to reduce this risk to an acceptable level.

Following a successful tender submission, the same process can be used to track and refine cost information to track progress and continue to highlight the potential risk areas. Subject to the constraints of the development process itself, it is feasible to re-allocate resource and re-order tasks in the process to reduce the risk in certain areas to bring a wayward project back on track. Larger organisations that have multiple projects running simultaneously can utilise the information provided by the tool to manage the risk and resource across their portfolio of projects.

A key feature of the tool is its ability to capture and utilise project duration data for use in providing more accurate estimates for future projects. The use of such corporate knowledge is particularly appropriate for organisations that produce variants of a product or undertake very similar projects. The number of organisations that may fully utilise this feature will depend greatly on the environment, and it may be most applicable for larger companies outside of New Zealand. To address this, the tool does not mandate the use of historical data therefore allowing it to be applied to both typical and atypical projects. For atypical projects, the underlying work breakdown structure can be modified to introduce new tasks for which historical data is not available and still produce a meaningful estimate for bespoke software produced by the small to medium sized enterprises that are typical of the New Zealand IT industry.

This paper first outlines the general environment in which the tool can be applied through a description of the software development lifecycle and a work breakdown structure for a typical project. Following an introduction to Monte-Carlo simulation and its applicability to uncertainty propagation the paper presents three scenarios for the use of the tool. The first scenario is the use of the tool to identify risk areas during a tender submission. The second scenario is the refinement of an effort estimate during the life of a project. Finally, the means of updating the historical database and refining the corporate knowledge base are shown.

**Project Environment**

*Software Lifecycle*

The software lifecycle is a term used to describe the various phases through which software travels. The software lifecycle runs from the point of conception to retirement. Whilst this paper assumes a waterfall style lifecycle, this is primarily as a means of aiding the clarity of presentation. The approach used to stochastically model the uncertainty can be applied to any lifecycle model, including the more agile approaches. The phases of the software development lifecycle include the traditional software development phases and the service management phases, combined into a single software lifecycle. The phases of the software lifecycle are:

- Concept
- Requirements Capture
- Analysis and Design
- Coding and Debugging
- Integration and Testing
- Deployment and Acceptance
- Maintenance
- Retirement and/or Replacement

The simple addition of features to existing software does not constitute the creation of new software and the beginning of a new software lifecycle. The software lifecycle of any software continues until it is formally ended by retirement.

Because the software lifecycle is cyclic, the same software can reside in different phases at the same time, requiring strong version, configuration and release control throughout the software lifecycle. While particular phases of the lifecycle may seem more significant than others, they are all crucial. All software must go through each phase of the software lifecycle at least once, and because of the circular nature of the software lifecycle, some more than once. The specific software design process used on a project is generally independent of the lifecycle model. This paper is not restricted to any particular software design process but assumes that a project work breakdown structure exists in line with the representation of a generic software lifecycle shown in Figure 1.

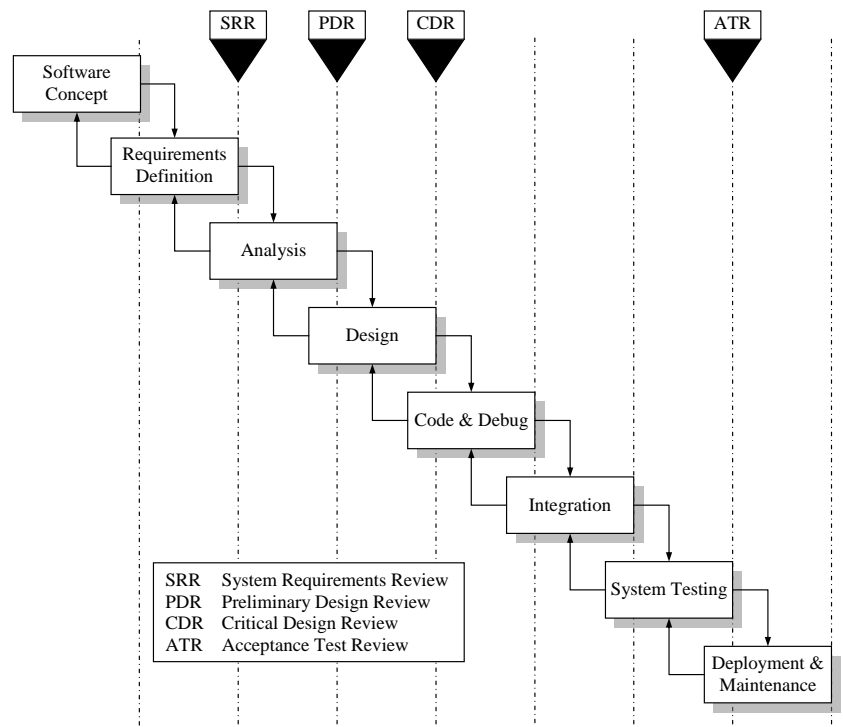

**Figure 1. The software lifecycle**

In the software development lifecycle, these phases should be considered as "super tasks". Super tasks are groupings of hierarchically ordered tasks and activities that need to be undertaken to achieve the goal of the super task. Such a task hierarchy is shown in Figure 2.

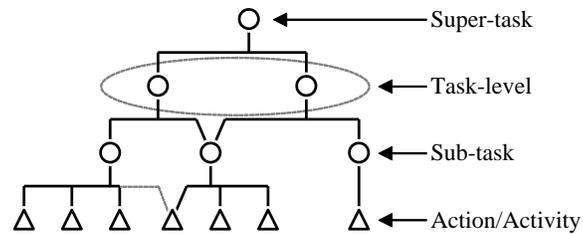

**Figure 2. Task Decomposition Tree**

Using the task decomposition tree in Figure 2, the relationship between the software development lifecycle, the software development process and the lower level activities becomes clear. Each phase of the lifecycle is a super task composed of tasks which are defined in the work breakdown structure. Again, each of these tasks can be decomposed into sub-tasks and activities. However, in this paper this decomposition is not undertaken and the work breakdown structure is kept purely at the task level. Activities and tasks that comprise the software design process are assumed to be conducted in the relevant super tasks, though this model can be modified to suit specific processes, work/product breakdown structures or software lifecycles as required.

*Work Breakdown Structure*

The work breakdown structure is shown in Figure 3, with work packages in normal type and project milestones in italic type.

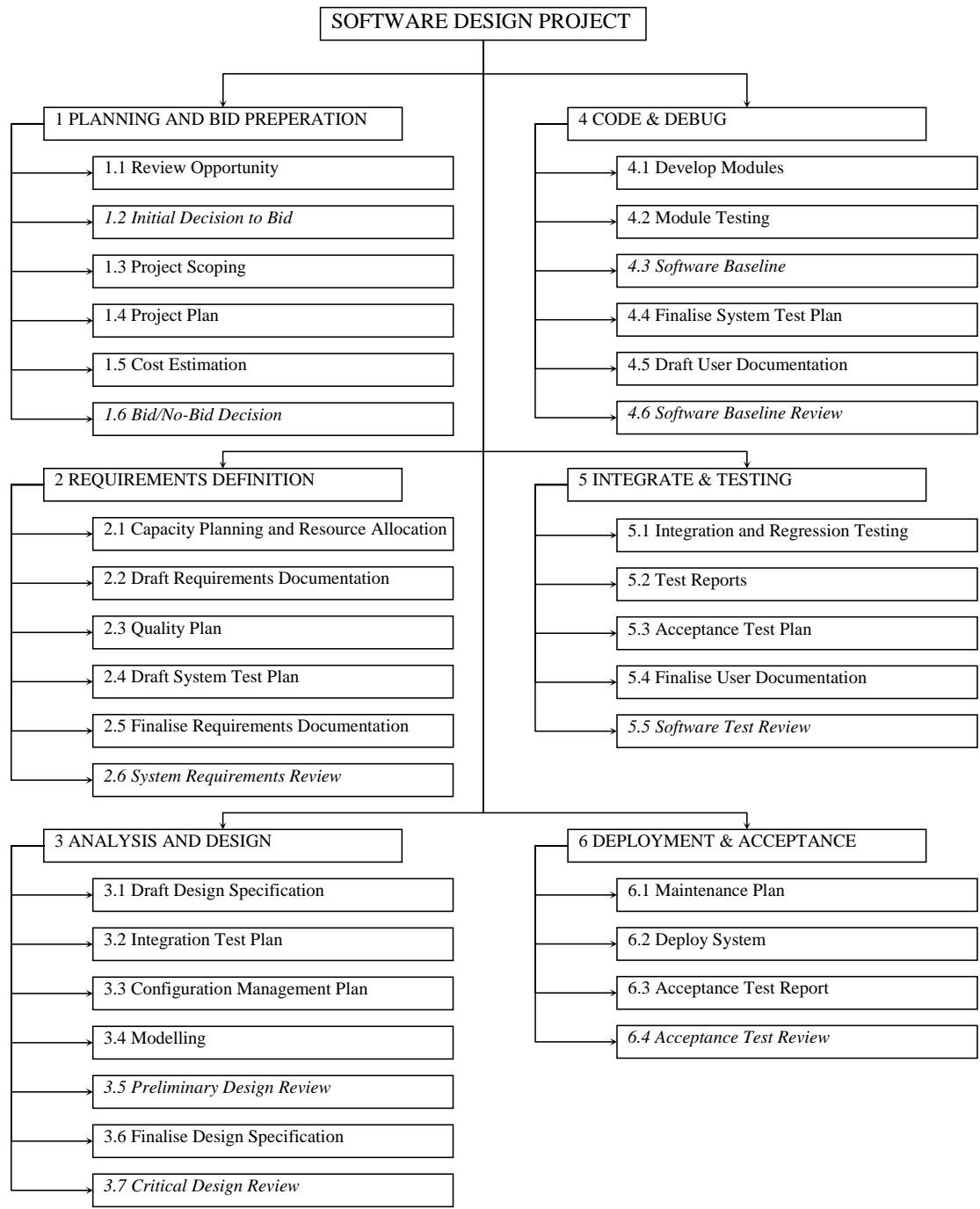

**Figure 3. Project Work Breakdown Structure**

The work packages in the work breakdown structure are in no way related to a specific design process, therefore actual day to day activities may be undertaken to satisfy more than one work package at any time. For example, in the bid preparation and planning phase, activities that support project scoping, the development of a project plan and a cost estimate will inevitably be conducted in parallel as there is co-dependence between tasks in each work package. However, in terms of the software lifecycle, the main reviews tend to be "gates" that limit a return to previous activities. For example, once the customer has approved the baseline design at the Critical Design Review then downstream activities will not include design unless it is at the customer request, which then is clearly a contractual change.

Each work package needs to be assigned an effort (and cost) estimate which can be developed using one or more of many methods available (see the following section). In the approach promoted here, each estimate can be defined using different probability distributions, namely a single value, a normal distribution, a triangular distribution and a uniform (rectangular) distribution. The choice of distribution and its corresponding parameters should represent the confidence in the estimate itself. Alternatively, a distribution may be selected on the basis of using historical data and fitting a distribution to that data. This approach accommodates a degree of corporate learning by utilising real outcomes of projects to aid the current project estimate.

**Effort and Cost Estimation**

In terms of new software development, it is not uncommon for effort or cost estimation to be done at the project concept (tendering) stage and for this single estimate to have a lifespan right through until the maintenance phase of the lifecycle, where the management model shifts towards bug fixes and enhancements which are treated as separate projects having their own cost/benefit analysis.

Estimates tend to be developed using a number of techniques, namely expert opinion, project analogy (use of historical data) or parametric models (Briand, El Emam, Surmann, Wieczorek, & Maxwell, 1999; Heemstra, 1990). In some cases, organisations will use a Pert estimate to combine estimates from different sources into a three-point estimate, with minimum, maximum and "most likely" cost estimates.

Whist this approach goes some way to mitigating risk in the cost estimation, there are two avenues that can be explored to further reduce risk. The first of these is the use of probabilistic modelling to gain a more realistic estimate of "most likely" cost. By assigning cost estimates against work breakdown structure items it is possible to use a Monte-Carlo simulation to provide a more realistic (and informative) estimate than that provided by a Pert estimate.

The second approach is to recognise that as a project matures so does the data that can be used in the cost estimation. During the concept phase, cost estimates against Work Breakdown Structure (WBS) items may simply be a wide range of values. As project tasks are undertaken, not only can these estimates be refined but the nature of the estimate can also be reconsidered. For example, it may be more appropriate to use a normal distribution, a three point (triangular) estimate or indeed even a point value. As the project further matures, completed WBS items would tend to be represented as single point values, further reducing uncertainty in downstream tasks.

The aim of this research is to develop a simple approach for cost and effort estimation that does not require the overhead of more formal approaches that include COCOMO-II (Boehm et al., 2000). However, the aim is not to replace such methods but augment them by providing additional tools. Monte-Carlo simulation provides a suitable means of introducing a powerful yet simple to use stochastic element to the cost estimation of software projects.

*Monte-Carlo Simulation*

A Monte-Carlo method is a technique that involves using random numbers and probability to solve problems using simulation. The approach has been used in a variety of problem domains, including cost estimation (Anderson & Cherwonik, 1997; Dimov & McKee, 1996). Computer simulation utilises computer models to imitate real life or make predictions. With a simple deterministic model a certain number of input parameters and a few equations that use those inputs produce a set of outputs, or response variables. A deterministic model, as shown in Figure 4, implies that the same results will be achieved no matter how many times the model is re-evaluated.

$x_1, x_2, x_3 \rightarrow$ Model $f(x) \rightarrow y_1, y_2$

**Figure 4. A parametric deterministic model**

Monte Carlo simulation is a method for iteratively evaluating a deterministic model using sets of random numbers as inputs. This method is often used when the model is complex, nonlinear, or involves more than just a few uncertain parameters. By using random inputs, the deterministic model is essentially transformed into a stochastic model.

The Monte Carlo method is just one of many methods for analysing uncertainty propagation, where the goal is to determine how random variation, lack of knowledge, or error affects the sensitivity, performance, or reliability of the system that is being modelled. Monte Carlo simulation is categorised as a sampling method because the inputs are randomly generated from probability distributions to simulate the process of sampling from an actual population. A distribution for the inputs that closely matches real data or best represents our current state of knowledge should be selected. If there is no specific data available, the best approach in the first instance is to apply a wide uniform distribution and allocate tasks that need to be undertaken to refine this estimate. The data generated from the simulation can be represented as probability distributions (or histograms) or converted to error bars, reliability predictions, tolerance zones, statistics and confidence intervals as illustrated in Figure 5.

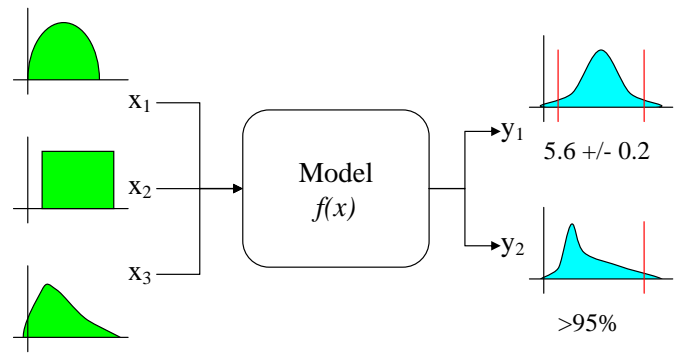

**Figure 5. Schematic showing the principle of stochastic uncertainty propagation**

The steps in Monte Carlo simulation corresponding to the uncertainty propagation are fairly simple, and can be easily implemented for simple models:

Step 1: Create a parametric model, $y = f(x_1, x_2, ..., x_q)$.
Step 2: Generate a set of random inputs, $xi_1, xi_2, ..., xi_q$.
Step 3: Evaluate the model and store the results as $y_i$.
Step 4: Repeat steps 2 and 3 for $i = 1$ to $n$.
Step 5: Analyze the results using histograms, summary statistics and confidence intervals

Monte Carlo simulation has been applied to modelling of uncertainty in cost estimations in a product breakdown structure (Crossland, Sims Williams, & McMahon, 2003) where historical project information is used to define the input probability distributions. This paper adopts a similar approach to the work breakdown structure representing the full life of a software project.

## Application of the Framework

*Minimising Risk in the Bid/No-Bid Decision*

Previous work (Barr, Burgess, Connor, & Clarkson, 2000) has developed a hierarchically structured model of the tendering process for technical domains. At the highest level this process model is entirely generic and can be applied to software development projects. Figure 6 shows this generic process model.

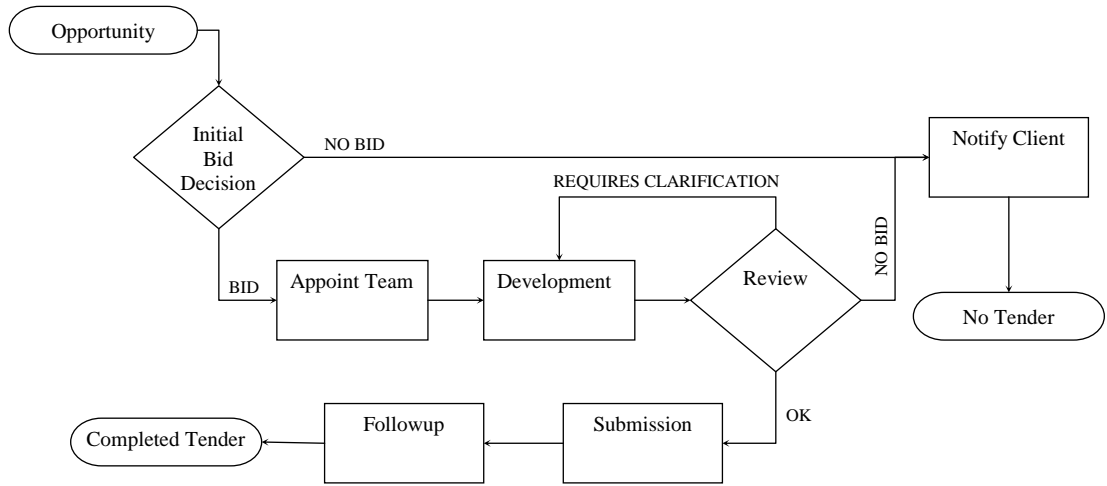

**Figure 6. Generic Tender Process Model (Barr, et al., 2000)**

The tool presented in this paper can be used in the initial response to an invitation to tender in order to gauge the risk in the proposed project and as such inform the bid/no-bid decision. In this application of the tool, it is assumed that minimisation of risk is conducted in the development activities conducted as part of the tender development process. The development activities can be decomposed into a specific lower level model that takes into account variations in the process between different domains and organisations. In this case the lower level model is defined by creating a process based on the work packages in the work breakdown structure in Figure 3. This lower level model is shown in Figure 7.

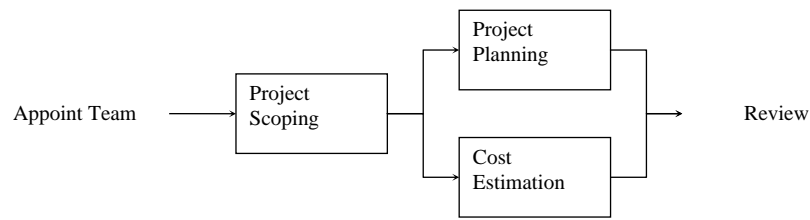

**Figure 7. Development Sub-Model (Barr, et al., 2000)**

In this model, a scope for the project is determined and a project plan and cost estimate is determined using additional lower level activities. These are not defined, but in the cost estimation area could include tasks such as "Obtain Expert Opinion", "Use Parametric Model" and "Analyse Historical Data". Iteration around the development activities occurs after the review of the data generated as illustrated in the generic top level model of tendering activities.

This approach to hierarchical modelling of processes is aligned with the hierarchical grouping of super tasks, tasks, sub tasks and activities presented in Figure 2. In this instance, the process and the software lifecycle are analogous. The process is composed of tasks, which at the highest level are common across many domains. These tasks are decomposed into sub-tasks and activities that are both domain and company specific.

This decomposition can in fact be continued to a lower level of granularity by relating activities to the artefact parameters that they manipulate. The current implementation is operating entirely in the process level but future work is intended to implement a signposting model (Clarkson, Connor, & Melo, 1999) of the software design process to allow the tasks in the process to be dynamically re-ordered on the basis of the confidence in the underlying parameter estimates.

In the tender process shown in Figure 6, the most crucial activity is the review of the tender documentation prior to submission. For many organisations, a poor review process with insufficient emphasis on identifying risks in the tender submission will result in a significant number of projects completed late or over budget. Lauesen and Vium (2004) have undertaken a study of typical problems identified in a competitive tendering process that can be used to assist in the identification of risk areas and future work will focus on the tailoring of the tool to address such risks. Applying the developed tool allows the risk in individual project phases to be quantified by using probability distributions to define the likely effort required to complete the phase. Figure 8 illustrates the means of entering this data into the tool.

| WBS Item | Type | P1 | P2 | P3 | Notes |
|---|---|---|---|---|---|
| **Planning & Bid Preperation** | | | | | Point Value |
| Review Opportunity (RO) | Point Value | 15.0 | | | P2 & P3 not used |
| Project Scoping (PS) | Historical | | | | |
| Project Plan (PP) | Historical | | | | Uniform |
| Cost Estimation (CE) | Historical | 30.0 | 5.0 | | P1 is min, P2 is max |
| | | | | | |
| **Requirements Definition** | | | | | Triagular |
| Capacity Planning/Resource Allocation (CR) | Uniform | 30.0 | 100.0 | | P1 is min, P2 is peak, P3 is max |
| Draft Requirements Documents (DR) | Uniform | 50.0 | 80.0 | | |
| Quality Plan (QP) | Historical | | | | Normal |
| Draft System Test Plan (TP) | Triangular | 25.0 | 50.0 | 60.0 | P1 is mean, P2 is StDev |
| Finalise Requirements Documents (FR) | Uniform | 20.0 | 90.0 | | |
| | Point Value | | | | Historical |
| | Normal | | | | Parameters not used |
| **Analysis & Design** | Triangular | | | | |
| Draft Design Specification (DS) | Uniform | 60.0 | 10.0 | | |
| Integration Test Plan (IP) | Historical | 50.0 | 75.0 | 100.0 | |
| Configuration Management Plan (CP) | Triangular | 20.0 | 60.0 | 100.0 | |
| Modelling (FM) | Historical | | | | |
| Finalise Design Specification (FD) | Triangular | 30.0 | 50.0 | 110.0 | |

**Figure 8. Data input**

From Figure 8, it can be seen that only four types of distribution (Point Value, Normal, Triangular and Uniform) may be selected manually, with the fifth option to determine the distribution from historical data. When this fifth option is selected, a much wider range of potential distributions will be tested against data values and a choice made as to which type of distribution best approximates the "real" data. To date, this historical database has been populated with dummy data for development purposes to establish the potential value of the tool. Further research will address this limitation as discussed in the conclusions.

Once the input values have been set to their initial values, the Monte-Carlo simulation is initiated, typically for between 5000 and 10000 evaluations. In each evaluation, a sample is taken for each input distribution and the output determined. Following completion of the simulation, the results may be viewed within the tool. Figure 9 shows the raw results and the statistics for the total project.

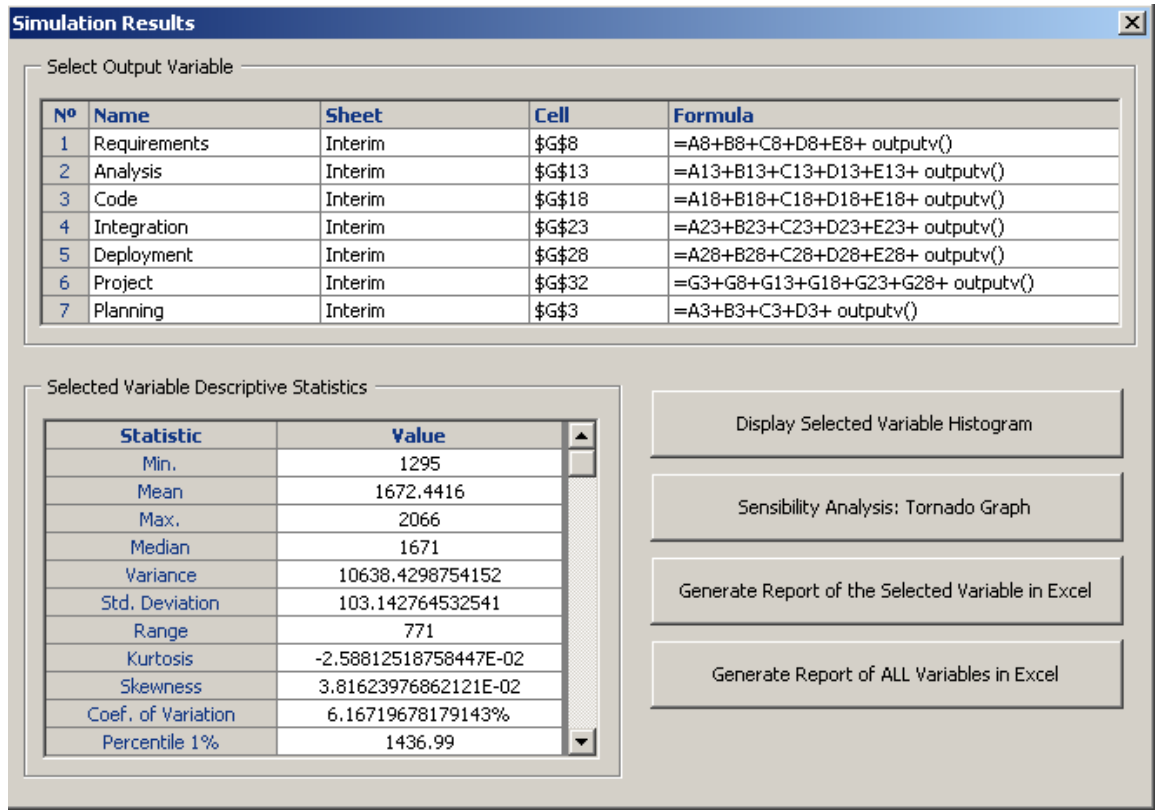

**Figure 9. Raw results**

The key statistics for considering the total project are the mean, the standard deviation and the interquartile range. Kurtosis and skewness are also important to consider but will be discussed in interpreting results from individual phases. Analysis of these statistics indicates that the simulation has predicted a wide range of outcomes that may constitute a project risk. Particularly the wide range of the duration and the large standard deviation are of concern. In addition to the statistics, the results for each output may be displayed graphically as a distribution of expected outcome. Figure 10 shows the expected outcome for the total project following completion of a simulation.

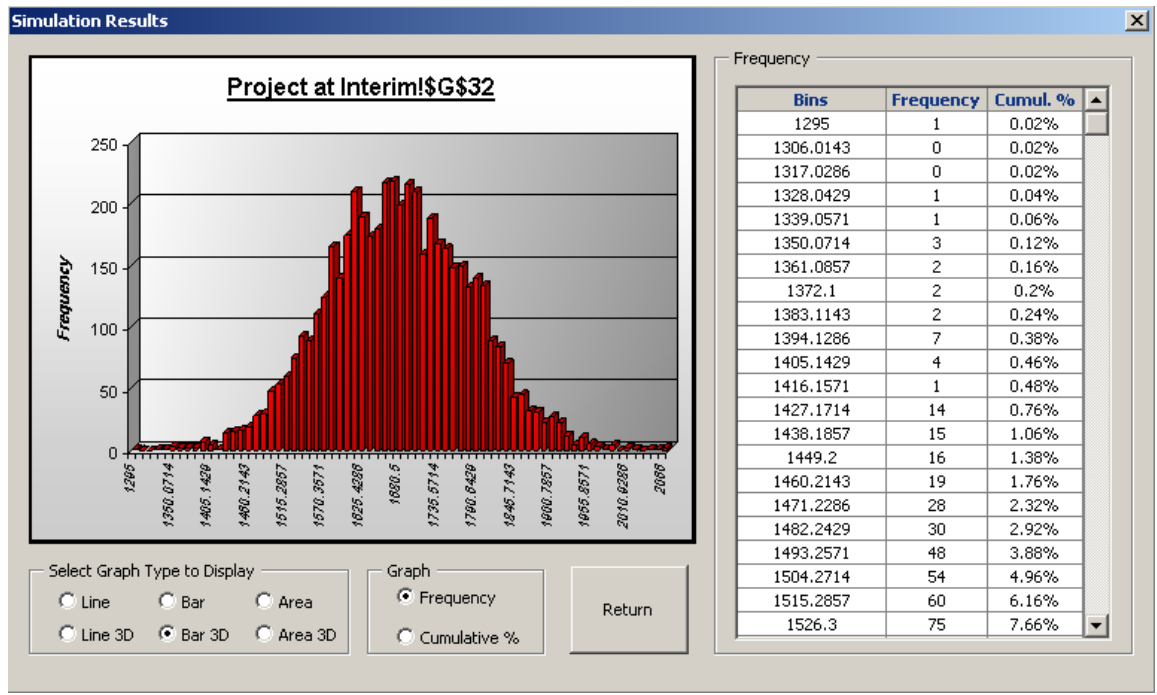

**Figure 10. Project duration distribution**

While an indication of likely duration for the entire project is useful, a more granular analysis could be even more informative. As has been demonstrated in previous work (Connor & MacDonell, 2005, 2006), the contribution of risk of each phase of the project to the total duration may be gauged by considering the distribution statistics for each phase. An indication of where the risks in the total project lie can be obtained by looking at the statistics associated with each individual phase of the project, particularly the Kurtosis, Skewness, Standard Deviation and the Interquartile Range. These statistics describe the shape and the spread of the distribution. This data can be plotted for each phase of the project to allow comparison to be made. For example, Figure 11 plots the Kurtosis of each phase such that the phase that is furthest away from the centre has the greatest risk.

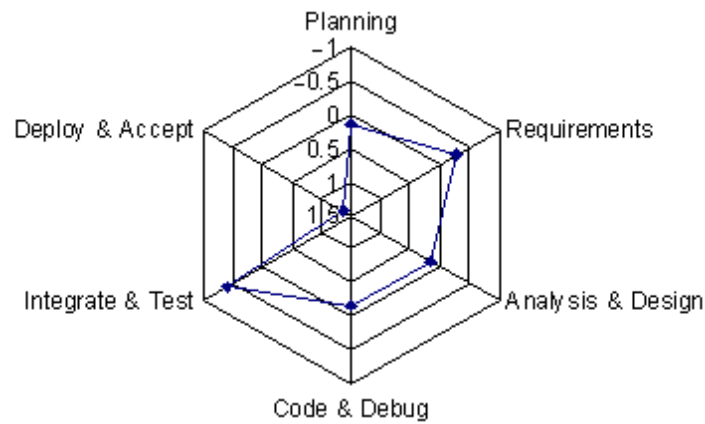

**Figure 11. Plot of kurtosis for each phase**

Project phases which exhibit a negative Kurtosis value have a more broad shape than a normal distribution, therefore the most negative value indicates a distribution that is tending towards being wide and flat. The nature of the distribution can be confirmed by plotting the results for this phase as in Figure 12.

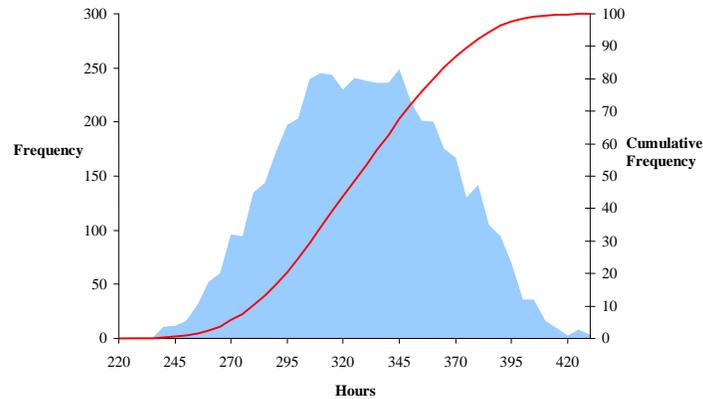

**Figure 12. Distribution of results for Integration and Test phase**

Using this metric, a refinement in the estimate for the Integrate and Test phase could result in an increased confidence in the overall project by producing an overall distribution with a more pronounced "spike", essentially implying a reduced level of risk.

Figure 13 plots the Skewness of each phase such that the phase that is furthest away from the centre has the greatest risk of overrun.

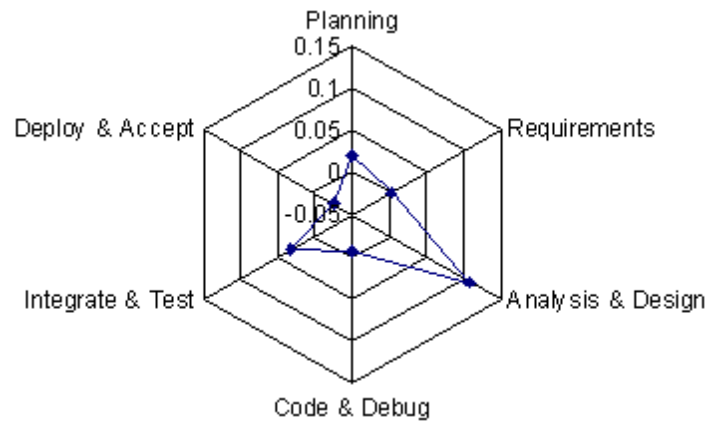

**Figure 13. Plot of Skewness for each phase**

Project phases which exhibit a positive Skewness value have a larger right tail than left tail, indicating that the phase is more likely to overrun than be completed early. Using this metric, a refinement in the estimate for the Analysis and Design phase could result in an increased confidence in the overall project by producing an overall distribution that is more centrally distributed or has a larger left tail, indicating likelihood to underrun. In managing projects, it is as important to identify underrun as to identify potential overruns. Underruns provide a degree of slack to compensate for overrun in either the project or the wider portfolio and can also be used to shift resource between tasks or projects.

*Estimate Refinement during Project Life*

In addition to the use of the tool in the tendering phase of a project, it has significant benefit in being used throughout the project life. To demonstrate this, the input settings of the example used above have been modified so as to represent a project in mid-life. Activities that have occurred in the past and are completed have been assigned point values. Activities that are towards the tail end of the project lifecycle can have their estimates refined as more knowledge is available on which to base the estimation. In this example, the project is assumed to be at the end of the requirements definition phase, so all activities in the planning and requirements phases have been set to point values. The activities in the Analysis and Design phase have been revised to be more precise and all other activity estimates have been untouched. Even these few changes have a significant effect on the overall project estimate as can be seen in Figure 14.

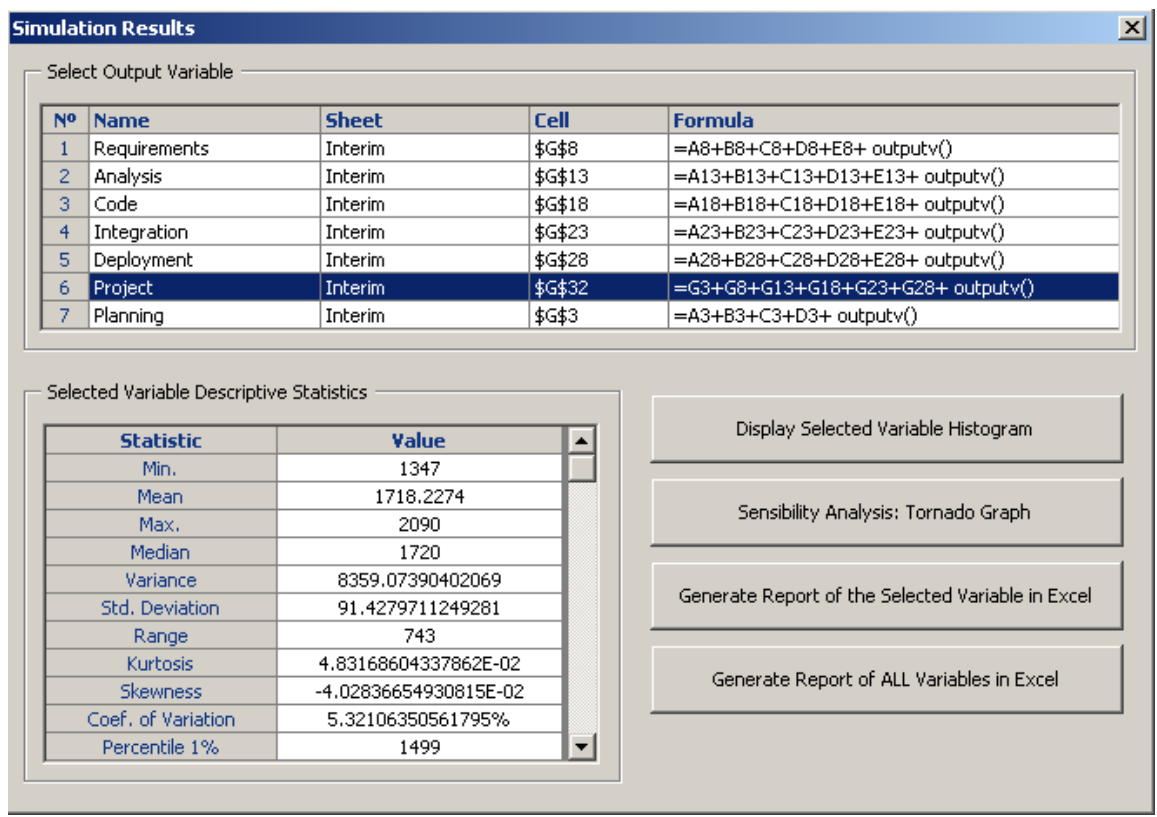

Figure 14. Revised simulation results

Whilst the mean estimate has increased, the standard deviation has reduced and, more significantly, both the kurtosis and the interquartile range have more favourable values. This shows that even a small change in confidence in the input parameters can result in a more realistic set of output distributions.

*Updating Historical Cost Database*

The use of a historical database provides a powerful tool for learning from previous experience and using this knowledge to inform future project estimates. The current implementation of the tool uses a simple means to capture and utilise historical data.

Historical data is captured within the Excel tool, simply as a list of actual effort required for each project broken down by project phase. The historical database is limited to typical projects, where typical is defined by the nature and scope such that they are within the expertise of the developers. The inclusion of atypical projects in the database does actually introduce an element of risk in the project estimates.

When new data is added to the database, it is necessary to refit a distribution to the data using the inbuilt functions of *SimulAr*. Figure 15 shows the original data set used for the Draft Requirements activity along with the best fit distribution. In this instance, the best distribution fit is achieved by using a logistic distribution and the quality of the fit is poor, as shown by the difference between the lines indicating the real data and theoretical distribution.

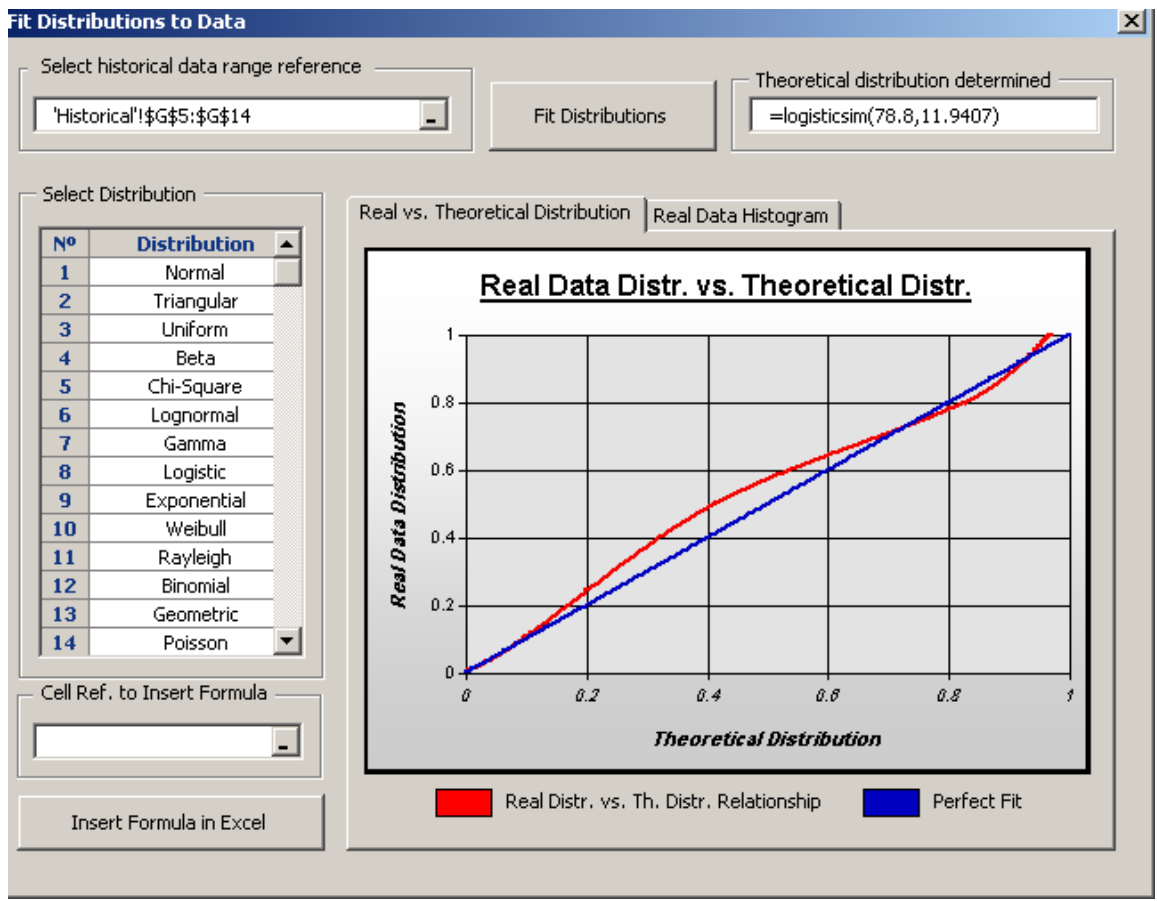

**Figure 15. Original fit of distribution to data**

Both the type and the value for the approximate distribution must be revised when new data is added. Even adding just one more entry into the database allows a higher quality of fit to be obtained, as illustrated in Figure 16.

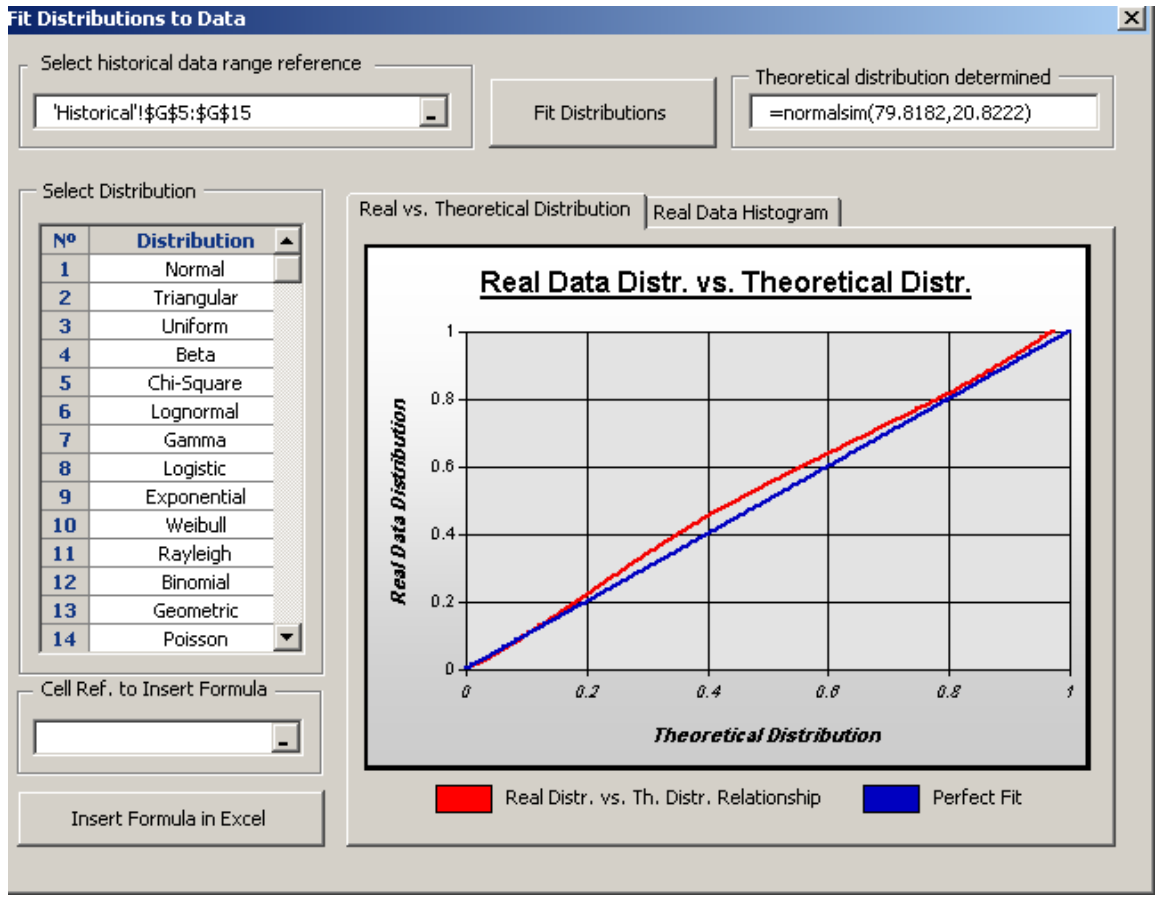

**Figure 16. Revised fit of distribution to data**

In this instance, a normal distribution provides the best fit to the actual data using 11 data points. Over time, as the database expands, the quality of fit will improve and the distributions become more representative of a typical project.

**Conclusions**

This paper has presented a methodology for tracking the uncertainty in project estimates and shown how modelling this uncertainty using probability distributions can inform both the submission of bids for projects and the subsequent project management itself. The software estimation process discussed in this paper describes the steps required for establishing initial software duration estimates and then tracking and refining those estimates throughout the life of the project. Establishment of this process early in the life cycle will result in greater accuracy and credibility of estimates and a clearer understanding of the factors that influence software development costs.

By linking estimates to a historical database of real project data, the approach has the capability to make accurate estimates early in the lifecycle with relatively low risk, despite the fact that the project requirements may be incomplete or inaccurate. The data in the historical data base is the actual duration of previous projects, for which estimates would have been made in similar circumstances when requirements were incomplete. For each and every project, corporate knowledge can be enhanced by comparing estimates at intervals throughout the lifecycle with the final cost or duration data at the end of the project. Whilst the number of small to medium sized enterprises that will be able to utilise this feature is limited, initial results have shown that it may be applicable to larger organisations that are frequently producing similar software products. Prior to undertaking live case study research, confidence in the performance can be obtained by undertaking a systematic study using publicly available datasets.

Further research will apply this tool to two available software project datasets. The first is provided by the International Software Benchmarking Standards Group (ISBSG). This non-profit group collects metric data on software projects from all over the world (currently twenty countries are represented), their most recent release comprising data on more than 3000 projects. The second repository is commonly referred to as 'the Finnish data set'. It is the result of commercially driven initiatives by Software Technology Transfer Finland (STTF) and includes software projects from 1978 to 2004. In its current form the data set comprises 622 projects from more than 40 organisations representing various sectors including finance, public sector, manufacturing and telecommunications. The use of a tool for project data submission ensures standardisation of features included. Also the project data are carefully assessed at STTF by experts before being added to the data base (Maxwell & Forselius, 2000).

The use of such standardised datasets, split into training and verification subsets, will enable a number of areas to be explored in a controlled way. Firstly, it will enable the applicability of the approach to typical and atypical projects to be determined by systematically adding data from the datasets into the historical database and using the tool to predict outcomes for other projects in the verification subset. In addition to this, the datasets will allow a controlled comparison to be made with other estimation approaches.

Throughout this paper, reference has been made to the ability to use statistical information with regards the uncertainty propagation to inform the ordering and priority of project tasks. It is a challenge for future work to explore this concept further by developing more detailed process models and defining dependencies between tasks and how tasks relate to the underlying data that can be used to drive the dynamic ordering of the process.

CITATION: Connor, A.M. (2007) "Probabilistic estimation of software project duration", New Zealand Journal of Applied Computing & Information Technology, 11(1), 11-22